\documentstyle[preprint,prb,aps]{revtex}
\begin{document}
\draft
\title{Multiple-Scattering Theory of X-Ray Magnetic Circular Dichroism:\\
       Implementation and Results for Iron K-edge}
\author{Ch. Brouder\cite{byline}}
\address{Laboratoire de Min\'eralogie-Cristallographie, CNRS URA9,
  Universit\'es Paris 6 et 7, 4 place Jussieu, 75252 Paris Cedex 05,
  France.\\ LURE, b\^at. 209D, F91405 Orsay Cedex, France.}
\author{M. Alouani}
\address{Department of Physics,The Ohio State University\\
         Columbus Ohio 43210, USA.}
\author{K. H. Bennemann}
\address{Institut f\"ur Theoretische Physik, Arnimallee 14,
  D-14195 Berlin, Germany.}
\date{\today}
\maketitle
\begin{abstract}
An implementation of the multiple-scattering approach to
x-ray magnetic circular dichroism (XMCD) in K-edge x-ray absorption 
spectroscopy is
presented. The convergence problems due to the cluster size and the 
relativistic corrections are solved using an expansion of the
Dirac Green function for complex energies up to the second order in 1/$c$. 
The Fermi energy is dealt with via a complex plane integration.
Numerical methods used to obtain the semi-relativistic Green function in 
the whole complex plane are explained. We present a calculation of 
the magnetic circular
dichroism at the K-edge of bcc-iron including the core hole effect. 
A good agreement is found at high energy. The physical origins
of the XMCD spectrum near the edge and far from the edge are analyzed.
The influence of the core hole, the possibility of a multiple-scattering
expansion and the relation of XMCD with the spin-polarized density
of states are discussed. 
A simple interpretation of XMCD at the K-edge
is presented in terms of a rigid-band model.
\end{abstract}



\section{Introduction}
X-ray magnetic circular dichroism (XMCD) in a magnetic sample is the 
difference between absorption spectra obtained from right- and
left-circularly polarized x-rays. This experimental technique
was discovered in 1987, \cite{Schutz87} and gives information on the
magnetic contribution of each orbital and of each atomic species in a sample.
Still now, the detailed mechanisms that
give rise to XMCD at the K-edge are not entirely clear near
the edge and are unknown far from it. The purpose of the present paper
is to understand XMCD at the K-edge over the whole energy range used
in experiments. Our main tool is a new semi-relativistic equation
which is very useful from a computational point of view, and which
can be used to calculate other relativistic properties, such as
magnetic anisotropy.

In a previous paper, 
we described a preliminary approach to the calculation of the x-ray 
magnetic circular dichroism effect in x-ray absorption spectroscopy, 
within the framework of the multiple-scattering theory. \cite{Hikam} 
We  assumed that XMCD at the L$_{\mbox{\scriptsize{II,III}}}$-edges
of rare earths and $5d$ transition metals has a simple interpretation and
that XMCD at the K-edge was the most difficult case, because spin-orbit
acts on the photoelectron, which makes a physical picture difficult to build.
>From the experience that was gained in the intervening years, \cite{BB} we
know now that the L$_{\mbox{\scriptsize{II,III}}}$-edges of rare earths 
are more complicated than expected, and some progress has been made 
towards a reliable practical use of the experimental results
\cite{Pizzini94,StefaniaXAFS8} at the K-edge of transition metals. 

Although fully relativistic calculations of XMCD at the K-edge are numerous, 
\cite{Ebert88,Ebert88b,Ebert88c,Ebert91c,Ebert92b,Stahler93,%
Gotsis94,Strange90,Ebert93b,Ebert93,Gotsis94b,Lang94,Strange95,Ebert96} 
we believe that the use of a semi-relativistic approach
is still justified.
Its main advantage is that spin and space variables are uncoupled at 
the zeroth order. This enables us to use the full local cluster symmetry,
to build a clearer physical picture, and to make orientation averages.
The smaller scattering-matrix dimension of the semi-relativistic 
approach allows for the
calculation of bigger clusters and broader energy ranges. Finally,
the  existing multiple-scattering numerical programs can be easily 
extended to
calculate the  XMCD. For instance, Ankudinov and Rehr have adapted
their FEFF program \cite{Ankudinov} to calculate the extended structure in
XMCD at the L$_{\mbox{\scriptsize{II,III}}}$-edge of rare earths and 5d 
transition metals. They obtained good agreement for gadolinium.
A similar adaptation is possible, although more complex, for K-edge spectra.

Here we report on the calculation of the K-absorption spectra and the XMCD 
of Fe in the presence and absence of a static core hole in the 1$s$ state.
The influence of the core hole was taken into account by the
final-state rule which assumes that
the final state energies of the x-ray absorption process 
are measured in   the presence of a
static core hole, i.e.,  the dynamics of the excitation process are
neglected. \cite{barth82} This  rule has been used with success
to explain the satellite structures on the high energy side of the 
emission spectra of simple metals. \cite{barth82,al86} In particular,  
for Na and Al the L$_{\mbox{\scriptsize{II,III}}}$ emission satellites 
on the high-energy side of the main lines due to the double ionization 
of the 2$p$ core level is well reproduced by the calculations if the 
excited atom is treated as an impurity and the static electron-hole 
interaction is included in the
calculation. \cite{barth82,al86} The final state rule was also used to
calculate the L$_{\mbox{\scriptsize{III}}}$-edge of 3$d$ transition metal 
ferromagnetics,
\cite{alouani94} and the effect of the core hole is found to be very
important.
The determination of the self-consistent potentials
is carried out  within the  local density approximation with a linear 
muffin-tin orbital (LMTO) basis-set. \cite{andersen}  We use a supercell
geometry and treat the excited atom with a core hole as a single impurity
atom in a lattice. A 1$s$ electron is added to the conduction
states in the supercell, and we  let the system of N+1 electrons relax
selfconsistently. We have found that the photo-electron localizes on
the excited atom providing an efficient screening of the core
hole. Like the case of L$_{\mbox{\scriptsize{II,III}}}$ 
edge\cite{alouani94} up to  80\% of the 
photoelectron polarizes as a minority spin. Thus, the electronic structure
of the excited Fe atom is very close to that of a cobalt metal.
However, in this calculation we have found that the effect of the core
hole on the K-absorption edge and the XMCD is small, and this is because
the $p$-states of Fe are less affected by the presence of the core hole. 

In this paper, the determination of the
absorption spectra and the XMCD is carried out within the multiple scattering
theory.  During  the implementation of our early formalism
we encountered several difficulties that convinced us that the naive
approach used in Ref.~\onlinecite{Hikam} to take spin-orbit effect into
account is mathematically not sound. In the present paper, 
we use a different mathematical framework for the calculation of XMCD,
that can be
generalized to take relativistic effects in the valence $d$-shell for
L$_{\mbox{\scriptsize{II,III}}}$-edges.
This approach gives a direct expansion of the Dirac Green function
in powers of $1\over c$, it overcomes the divergence of the 
Foldy-Wouthuysen transformation through the use of complex energies,
and it leads to results in reasonable agreement with experiment.

Some of our results are useful only for x-ray absorption spectroscopy,
but others can have wider applications, such as the convergence of the
semi-relativistic limit and the properties of the Green function at
complex energies.

The remainder of the paper is organized as follows. In the second section
we point out the limitation of the Foldy-Wouthuysen transformation
and give a new method for the determination of the relativistic
effects within the Green function approach.
We then apply the Green function framework to the calculation
of the XMCD signal.
In a third section the use of complex energies is 
discussed, and a complex plane integration is used to take the Fermi energy 
into account.
In the fourth section we present the 
numerical implementation that we used, and give the new XMCD cross-section 
formula at the K-edge.  Finally, we present a calculation that was
carried out on a reasonably large Fe cluster of 259 atoms near the
edge and a cluster of 51 atoms up to 500~eV. Good agreement with
experiment is found in the high-energy region.
Various issues, such as the influence of the relativistic core hole,
the origin of XMCD at the edge and far from it, the possibility
of a multiple-scattering expansion are discussed. The spectra
are related to the density of states through a simple rule, and
the rigid-band model is found to be correct at high energy.
An appendix showing how symmetrized bases were used
to calculate XMCD faster and to understand some aspects of the
experimental spectra closes the paper.

\section{The relativistic expansion}
\subsection{The Foldy-Wouthuysen transformation}
The relativistic corrections to physical phenomena are usually
treated using a perturbation approach developed in 1950 by Foldy
 and Wouthuysen (FW), \cite{Foldy} who
made a unitary transformation to eliminate the coupling of the large
small components from
the Dirac equation and obtained, to zeroth order, the Pauli Hamiltonian
and, to first order in ($1/c^2$), the spin-orbit, Darwin and kinetic energy 
corrections.
Later, this Hamiltonian was taken as a starting point for the calculation
of many physical properties, such as magnetic anisotropy, XMCD or 
magneto-optic effects. Sometimes, the spin-orbit term
is used in a second-order perturbation theory, overlooking that the
FW-transformation must then also be made up to second
order ($1/c^4$) to achieve formal consistency. 
However, when carrying out higher-order transformations,
strongly divergent terms are found even for the simple case of
a hydrogen atom: \cite{Morrison}  the second order ``correction'' is
a sum of infinite terms. Additional information concerning the 
convergence properties of the FW-transformation can be found in 
Ref.~\onlinecite{Gesztesy}.

Even at the $1/c^2$ order, the behavior of the
Hamiltonian obtained by adding the kinetic energy correction
to the non-relativistic Hamiltonian
is so bad that all negative eigenvalues (bound states) disappear,
and a continuum of states is obtained between $-\infty$ and $mc^2/4$,
however small $c$ may be. \cite{Gesztesy85} In some cases, 
\cite{Hikam,Nagano} only
the spin-orbit interaction is added to the non-relativistic Hamiltonian.
For a Coulomb potential, the negative spin-orbit term pulls the
particle into the singularity and the resulting Hamiltonian is
not essentially self-adjoint (it can have any eigenvalue, depending
on the chosen self-adjoint extension). 
\cite{GesztesyPL,Udim,ThallerPC} 
Physically reasonable results could be obtained by choosing the boundary 
condition where the wave function is zero at $r=0$ but this process 
is mathematically ambiguous. \cite{ThallerPC}

To summarize, the series obtained from the FW-transformation 
exhibits three kinds
of divergence: (i) the series itself diverges, which is usually not a
problem since most perturbation series used in quantum mechanics
are divergent (e.g. Zeeman effect, Stark effect, anharmonic oscillator);
\cite{Hunziker} (ii) from the second order term ($1/c^4$), each
term of the series is a sum of infinite terms that should add to
a finite term, but no procedure is known to carry out this 
summation;\cite{Morrison} (iii) the first order
term of the series (spin-orbit coupling) diverges as the cluster
size increases. 

Recently, all these problems of the semi-relativistic limit were
solved using modern mathematical methods, and we propose to use this new 
formulation for the calculation of XMCD. Other approaches to the 
semi-relativistic limit of the Dirac equation were proposed by 
quantum chemists. 
\cite{Kutzelnigg,Sadlej}

\subsection{Relativistic $(1/c)$ expansion of the Green function}
The standard method to calculate relativistic effects within the 
Green function approach is to consider the spin-orbit and other 
terms as a perturbation of the
Hamiltonian, and to use the Lippmann-Schwinger equation corresponding
to that perturbation. \cite{Hikam,Nagano} This procedure is not
mathematically safe and leads to divergences.  Another approach to XMCD
was proposed by Natoli \cite{RinoXAFS8} where the Schr\"odinger 
equation including spin-orbit was solved for each muffin-tin sphere.

Here, we start from the retarded Green function $G^D(z)$ corresponding
to the Dirac Hamiltonian with potential $V$ (the Dirac Green function)
and we use a slight modification of 
the analytic $1/c$ expansion of $G^D(z)$ that was obtained
in Ref.~\onlinecite{Gesztesy} (see Ref.~\onlinecite{brouderMittelwihr}
for a detailed proof):
\begin{eqnarray}
G^D(z)&=&{[1-T(z)]}^{-1}
\left(\begin{array}{cc} G(z) & {G(z)A^{\dag}/ (2mc)}\\
{AG(z)/(2mc)} & {[AG(z)A^{\dag}+2m]/(2mc)^2}\end{array}\right)
\label{NRExp}
\end{eqnarray}
where
\begin{equation}
T(z)=\left(\begin{array}{cc} 0 &  {G(z)A^{\dag}(V-z)/(2mc)}\\
 0 & {[AG(z)A^{\dag}+2m](V-z)/(2mc)^2}\end{array}\right).
\end{equation}

All matrix entries are 2x2 matrices.
$A=-i\hbar{\bf\sigma.\nabla}$ and ${\bf\sigma}$ are the Pauli matrices
and $G(z)$ is the Green function for the Pauli Hamiltonian with 
potential $V$. $V$ is generally a 2x2 matrix that describes the
potential experienced by the electrons with up and down spins.
In the simplest case, $V$ is a diagonal matrix made up of 
$V^\uparrow$ and $V^\downarrow$.
The successive terms of the relativistic expansion are obtained by the 
series ${[1-T(z)]}^{-1}=1+T(z)+T^2(z)+\cdots$.
This expansion gives a very compact formulation of the
correction terms. For instance, the first order correction to
a non-degenerate bound-state eigenvalue $E_0$ corresponding to the
eigenstate $|\psi\rangle$ is given by
$$
\Delta E=\langle\psi|A^{\dag}(V-E_0)A|\psi\rangle,
$$
\cite{Thaller}
which contains the kinetic energy correction, the Darwin correction and
the spin-orbit correction. Because of this simplicity, formal manipulation
can be carried out much further, for instance to obtain a
consistent second order expansion of the spin-orbit interaction.
Another aspect of this expansion is that the small components are not
eliminated, which will be useful for x-ray absorption spectroscopy.

It was shown that expansion (\ref{NRExp}) is analytic, \cite{Gesztesy} but
its actual radius of convergence  was investigated only recently.
Preliminary results\cite{White,Arai}
for the Dirac Green function without potential
showed that the radius of convergence was a function of the imaginary
part of the energy. For a Dirac equation with potential, 
the Green function expansion could be shown \cite{White2} to be 
convergent only
when the imaginary part of the energy is greater than $mc^2$
(a very bad resolution indeed). However, the
convergence can be much better when a physical property is calculated
instead of the general Green function. For instance, the convergence
of the bound state energy expansion is fast. \cite{White,Morin}
Physically, the idea that can be drawn from these mathematical results
is that the use of complex energies makes the semi-relativistic expansion
converge, whereas it diverges on the real axis. This idea will be
detailed in the next section.

>From the formal point of view
 fully relativistic programs require a coupling
of an infinity of different $l$ values due to the magnetic field.
\cite{Strange} Up to now, the coupling between $l$ and
$l\pm 2$ was neglected, which limits the maximum possible value of $l$.
\cite{Strange2} In our approach, this coupling is made consistently
with $1/c$ through the expansion process.

\subsection{Application to XMCD \label{SectXMCD}}

In x-ray
absorption spectroscopy, the spin-orbit parameter for the core hole
has a magnitude of several hundreds of eV and cannot be considered 
{\it a priori} small.
Furthermore,  relativistic effects in x-ray absorption have been shown 
to be considerable both theoretically \cite{Trevor} and experimentally. 
\cite{Jeon}
Therefore, a fully relativistic core hole wave function must be used.
This is not possible within the standard semi-relativistic approach. 
In the electric dipole approximation, the relativistic formula
for the x-ray absorption cross-section is:
\begin{equation}
\sigma(\hat\epsilon)=-4\pi\alpha_0\hbar\omega\langle i|
   (\hat\epsilon^*.{\bf r})\,{\rm Im}[G^D({\bf r},{\bf r}';z)]
   (\hat\epsilon.{\bf r}')|i\rangle,
\end{equation}
where $z=\hbar\omega+E_i+i\epsilon$, $E_i$ is the energy of the
core level $|i\rangle$ and $\hat\epsilon$ is the x-ray polarisation
vector.
The calculation of the fully relativistic initial state $|i\rangle$
presents no difficulty, thanks to the availability of Desclaux' program.
\cite{Desclaux} 
In Ref.~\onlinecite{Hikam} we supposed that
the relativistic effects on the valence and continuum
states are weak. We show in the present section that this supposition is
not always true. 

The influence of relativistic effects on the photoelectron is obtained
by the series expansion of the Dirac Green function.
The Dirac Green function in Eq.(\ref{NRExp}) is expanded up to second 
order in $1/c$: \cite{GesztesyPRL}

\begin{eqnarray}
\langle i|{\bf \hat\epsilon^*.r}G^D{\bf \hat\epsilon.r'}|i\rangle&\simeq&
\langle \phi|{\bf \hat\epsilon^*.r}G{\bf \hat\epsilon.r'}|\phi\rangle 
   \nonumber\\
&+& {1\over 2mc}\langle \phi|{\bf \hat\epsilon^*.r}GA^{\dag}
        {\bf \hat\epsilon.r'}|\psi\rangle
+ {1\over 2mc}\langle \psi|{\bf \hat\epsilon^*.r}AG
        {\bf \hat\epsilon.r'}|\phi\rangle
   \nonumber\\
&+& {1\over (2mc)^2}\langle \phi|{\bf \hat\epsilon^*.r}GA^{\dag} (V-z)AG
        {\bf \hat\epsilon.r'}|\phi\rangle
   \nonumber\\
&+& {1\over (2mc)^2}\langle \psi|{\bf \hat\epsilon^*.r}(2m+AGA^{\dag})
        {\bf \hat\epsilon.r'}|\psi\rangle \label{NonRelExp}
\end{eqnarray}
where  the two-component spinors $|\phi\rangle$ and $|\psi\rangle$ 
are the large and small components of the Dirac spinor $|i\rangle$.
In our previous approach \cite{Hikam} we have neglected the relativistic 
nature of the core state, and used only the spin-orbit interaction, as a
consequence we had obtained only the first and a part of the fourth term.
Formula (\ref{NonRelExp}) is valid for a non-muffin-tin potential;
the use of non-muffin-tin potentials is probably required
to achieve a quantitative estimate of XMCD.
In all  the terms of Eq.~(\ref{NonRelExp}) only the fourth term
yields magnetic circular dichroism for a powder sample. 
Therefore, in the following,
we call $\sigma_0$ the spectrum obtained from the first term,

\begin{equation}
\sigma_0(\hat\epsilon)=-4\pi\alpha_0\hbar\omega\sum_s\langle \phi^s|
   (\hat\epsilon^*.{\bf r})\,{\rm Im}[G({\bf r},{\bf r}';z)]
   (\hat\epsilon.{\bf r}')|\phi^s\rangle
\end{equation}
where the sum over $s$ is the sum over initial states (two spin
states for a K-edge). For a powder, $\sigma_0$ does not depend on
$\hat\epsilon$. We call $\sigma_1$ the spectrum obtained from
the fourth term of Eq.~(\ref{NonRelExp}).

\begin{equation}
\sigma_1(\hat\epsilon)=-4\pi\alpha_0\hbar\omega
  \sum_s\langle \phi^s|(\hat\epsilon^*.{\bf r})\,
  {\rm Im}\left[G(z){A^{\dag}(V-z)A\over(2mc)^2}G(z)\right]
   (\hat\epsilon.{\bf r}')|\phi^s\rangle \label{sigma1}
\end{equation}
and $\sigma_{MCD}=\sigma_1(\hat\epsilon^-)-\sigma_1(\hat\epsilon^+)$ 
for an external magnetic field aligned with the x-ray wavevector.

More precisely, we have in the fourth term: \cite{Thaller}
\begin{equation}
A^{\dag}(V-z)A=-\hbar^2[(\nabla V.\nabla)+(V-z)\Delta+
i\sigma.(\nabla V\times\nabla)].
\end{equation}
In the right-hand side of the above equation, only the third term 
contributes to XMCD at the K-edge, because the first
two terms do not connect space and spin variables.
In the case of spherical potentials, one obtains the usual formula for
the spin-orbit interaction: \cite{Cowan}
\begin{equation}
\xi(r)=-i{\hbar^2\over (2mc)^2} \sigma.(\nabla V\times\nabla)=
{\alpha_0^2 a_0^2\over 4} {1\over r}{dV\over dr} \ell.\sigma.
\label{SOEq}
\end{equation}
We took this formula for convenience, to make contact with 
Ref.~\onlinecite{Hikam}, 
but it is also possible to act with $A$ and $A^{\dag}$ directly on the Green
function. This would be much simpler if all relativistic effects
(and not only XMCD) were investigated. \cite{Trevor,Jeon} In principle,
$\sigma_1(\hat\epsilon^-)+\sigma_1(\hat\epsilon^+)$ contributes also
to the normal absorption spectrum, but this relativistic contribution
was assumed to be small as compared to $\sigma_0$.

Finally, to obtain the influence of all sites, the potential $V$ must
be written as a sum over all sites plus an interstitial potential. 
Because the operator $A$ is translation
invariant, each site can be taken as the origin when making
transformation
(\ref{SOEq}). In the muffin-tin case, it can be shown that the 
interstitial region does not contribute to XMCD.

For real energies, XMCD calculated within the present approach is formally
identical with the results of Ref.~\onlinecite{Hikam}. Therefore, we can use
the latter to understand the real energy behavior of the former.
For small clusters, the calculation of XMCD was easy, \cite{HikamTh} 
but as the cluster size increased, we met difficulties. 
Fig.~\ref{FebccEi0} shows 
$\sigma_0$ and $\sigma_{MCD}$ for a cluster of 259 atoms of bcc-Fe.
Both spectra exhibit a very sharp resonance
(width $\approx$ 0.005~eV). This resonance is due to the fact that, 
as the cluster
size increases, one comes close to the singularities of the crystal
scattering matrix. \cite{Vedrinskiy} The second point is that $\sigma_{MCD}$ 
is larger than $\sigma_0$.
This unphysical result proves that the spin-orbit effect is not a small
perturbation of the system in that energy range. This illustrates the
fact, discussed in the previous section, that the relativistic expansion
is generally not convergent at real energies. It may be worthwhile 
noticing that this divergence of the relativistic expansion is not
due to the $Z/r^3$ singularity of the spin-orbit interaction in a 
Coulomb potential.

Moreover, the experimental resolution at the K-edge of transition metals is
of the order of 1~eV whereas Fig.~\ref{FebccEi0} shows much sharper 
structure. 
Therefore, the usual method of convoluting the theoretical spectrum 
with a Lorentzian profile for 
comparison with experiment would mean calculating much more 
points than actually needed.

For real potentials, the singularities of the multiple-scattering matrix
are on the real axis, and the use of a complex energy smoothes the sharp
structure of $\sigma_0$ linked with the cluster size. 
We show in the next section that the convergence problem of the
relativistic expansion is also solved by calculating the Green function 
at complex energies.

Finally, we work within a real space multiple-scattering approach, so that
the cluster is assumed to be of finite size. The relativistic expansion
of the Green function for an infinite crystal was studied in 
Ref.~\onlinecite{Bulla}.

\section{Complex energies}

In this section, we justify the use of complex energies, and we
show how this simplifies the calculation
of physical quantities. We assume that the potential
$V$ is a diagonal 2x2 matrix made up of $V^\uparrow$ and $V^\downarrow$,
which are the potentials experienced by up and down spins. These
potentials are assumed to be real. The generalization to complex
potentials is a non-trivial task,\cite{Agmon} and the use of an
energy-dependent width $\Gamma(E)$, which can be considered to be the
imaginary part of the potential (constant over space), was
found to be sufficient to reproduce experimental 
spectra. \cite{Sainctavit}

\subsection{Green function}
The link between the wave function and the Green function 
formulae for x-ray absorption cross-section is established 
through the identity:\cite{Newton}
\begin{equation}
\sum_f|f\rangle\langle f|\delta(E-E_f)=\delta(E-H)=
-{1\over 2\pi i}[G^+(E)-G^-(E)],
\end{equation}
where $G^\pm(E)=G(E\pm i\epsilon)$,
$\epsilon$ is a positive number that is taken to tend to zero at
the end of the calculation (this is just a notation to designate
an integration path in the complex plane).

Because of the finite core-hole lifetime, the calculated spectrum
must be convoluted by a Lorentzian with a half width at half maximum
(HWHM) $\Gamma$ (which
may ultimately depend on the photoelectron energy).
In the Green function formalism, the convolution with a Lorentzian
is obtained by calculating the Green function for a complex energy. 
\cite{Rino}  This can be shown from the formula:
\begin{equation}
\int_{-\infty}^{+\infty}de{\Gamma\over\pi}
{G^+(e)-G^-(e)\over (E-e)^2+\Gamma^2}=
G(E+i\Gamma)-G(E-i\Gamma)\label{IntLor}.
\end{equation}
Since $G({\bf r},{\bf r}';z)=G({\bf r}',{\bf r};z)$ \cite{Agmon}
and $G({\bf r},{\bf r}';z)^*=G({\bf r}',{\bf r};z^*)$ for Hermitian
potentials $V$, we have $G(E-i\Gamma)=G(E+i\Gamma)^*$.

Moreover, the term $\hbar\omega=e-E_i$ 
of the absorption cross-section can
be taken care of by noticing that $zG(z)=HG(z)+1$.
Thus $e G^\pm(e)=HG^\pm(e)+1$, which gives, after convolution, 
$HG(E+i\Gamma)+1=(E+i\Gamma)G(E+i\Gamma)$. 

Therefore
\begin{equation}
\sigma_0(\hat\epsilon)=-4\pi\alpha_0\langle i|(\hat\epsilon^*.{\bf r})
   \,{\rm Im}[(E+i\Gamma-E_i)G({\bf r},{\bf r}';E+i\Gamma)]
    (\hat\epsilon.{\bf r}')|i\rangle.
\end{equation}

In other words, the convoluted spectrum is obtained by calculating
the Green function for an energy with an imaginary part $\Gamma$.

\subsection{Fermi energy}
In the previous section, we did not take into account  that
all one-electron states up to the Fermi energy $E_F$ are occupied.
To yield physical results, integral (\ref{IntLor}) must be carried out from
$E_F$ instead of from $-\infty$.
This modifies our expression in an interesting way. We must now
evaluate :

\begin{eqnarray}
&&{1\over 2i}\int_{E_F}^{+\infty}de{\Gamma\over\pi}
{(e-E_i)(G^+(e)-G^-(e))\over (E-e)^2+\Gamma^2}=\nonumber\\
&&{1\over 4\pi}\int_{E_F}^{+\infty}de (e-E_i)(G^+(e)-G^-(e))
\left[{1\over E-e+i\Gamma}-{1\over E-e-i\Gamma}\right ]. \label{EfInfty}
\end{eqnarray}

This integral can be calculated by a complex plane integration technique:
\cite{Dreysse}
Since the self-consistent potential $V$ is Hermitian, the poles of
$G^+(e)$ are at $e=r-i\epsilon$, where $r$ is a real number.
Therefore, we can choose the contour of Fig.~\ref{poles} to
apply Cauchy's integral formula.\cite{Byron} 
The contribution of the Jordan contour at infinity is zero \cite{Jordan} and
we obtain:

\begin{eqnarray}
&&{1\over 2i}\int_{E_F}^{+\infty}de{\Gamma\over\pi}
{(e-E_i)G^+(e)\over (E-e)^2+\Gamma^2}=-{i\over 2}\theta(E-E_F)
   (E+i\Gamma-E_i)G(E+i\Gamma)\nonumber\\
&&+{\Gamma\over\pi}\int_{0}^{\infty}dt
{(E_F+it-E_i)G(E_F+it)\over (E_F+it-E)^2+\Gamma^2}.
\end{eqnarray}
Heavyside's step function $\theta(E-E_F)$ is present in the expression
because, when $E>E_F$, the pole $E+i\Gamma$ is inside the contour.

The second integral in the rhs of Eq.~(\ref{EfInfty}) is closed 
on the negative imaginary side, and we obtain finally:

\begin{eqnarray}
&&\int_{E_F}^{+\infty}de{\Gamma\over\pi}
{(e-E_i){\rm Im}[G^+(e)]\over (E-e)^2+\Gamma^2}=\theta(E-E_F)
   {\rm Im}[(E+i\Gamma-E_i)G(E+i\Gamma)]\nonumber\\
&&+{\Gamma\over\pi}\int_{0}^{\infty}dt
{\rm Re}\left[{(E_F+it-E_i)G(E_F+it)\over (E_F+it-E)^2+\Gamma^2}\right].
\label{FermiInt}
\end{eqnarray}

It can be shown that the right-hand side of Eq.~(\ref{FermiInt})
is continuous at $E=E_F$ in spite of the step function. \cite{Continuous} 
The recourse to this contour integration is efficient, as compared to
an integration over real energies, because the Green function is quite
smooth on the line $z=E_F+it$ (see Fig.~\ref{G(EF+it)}) and only a few
points must be calculated to evaluate the integral. Moreover, 
$G(E_F+it)$ tends rapidly to the local atomic Green function of the absorbing
site, as shown in Fig.~\ref{G(EF+it)}. Therefore, $G(E_F+it)$ can be 
obtained at
high $t$ just by calculating a simple one-site Green function. The 
vanishing of the neighbor's influence can be understood through the form 
of the structure constant matrix $H$ (see section \ref{NumerSec}).
The $H$-matrix
elements have the form $H^{ij}=\exp{(i\kappa R_{ij})}P(1/\kappa R_{ij})$
where $P$ is a polynomial.
For large $t$,  $\kappa=\sqrt{z}\simeq\sqrt{t/2}(1+i)$, and the $H$-matrix
elements describing the influence of the neighbors are damped by a 
factor of $\exp{-\sqrt{t/2}R_{ij}}$. The integration in the complex plane
was chosen along a straight line. This is not a steepest descent path but
was found to be sufficiently efficient.

The above proofs are the same whether $G(z)$ is the non-relativistic or
the relativistic Green function (assuming that the negative energy
states are full). To obtain the expression for $\sigma_1$ after
convolution, we start from the
results obtained with the Dirac Green function and use the
relativistic expansion (\ref{NonRelExp}). This gives an expression
for $\sigma_1$ which is the sum of Eq.~(\ref{sigma1}) for $z=E+i\Gamma$
and of an integral over the line $z=E_F+it$. The detailed form of
the result will be given in section~\ref{crosssection}.

It was observed by Rehr \cite{Rehr} that thermal motion can also 
be a convergence
factor in multiple-scattering calculations. Although thermal effects
can be formally accounted for in full multiple-scattering calculations,
\cite{Boston} temperature effects were neglected in the present study.

Finally, a characteristic of the $1/c^2$ relativistic term is
that spin-flip is not allowed for powders. An explicit calculation
of the spin-flip scattering amplitude shows that it is an order
of magnitude smaller than non-spin-flip scattering for Nd$^{3+}$.
\cite{Zhogov95} Experiments \cite{Mulhollan} confirm that
elastically scattered electrons are rarely spin-flipped.

\section{Numerical aspects}
\label{NumerSec}

The x-ray absorption cross-section is  obtained from the cluster
Green function for complex energies, which  we write as 
\cite{Gyorffy,brouderMittelwihr}
\begin{eqnarray}
&&G({\bf r}_i,{\bf r}'_j;z)=-i\kappa t^i_\ell\sum_{\ell m}
        {R^i_\ell(r_<)Y_\ell^m(\hat r_i) \over \sin\delta^i_\ell}
        H^i_\ell(r_>){Y_\ell^m}^*(\hat r_i')\delta_{i,j} +\nonumber\\
&&\kappa^2\sum_{\ell m\ell'm'}
      {R^i_\ell(r_i)Y_\ell^m(\hat r_i) \over \sin\delta^i_\ell}
      (\tau^{ij}_{\ell m\ell'm'}
     +{t^i_\ell \over \kappa}\delta_{\ell,\ell'}\delta_{m,m'}\delta_{i,j})
    {R^j_{\ell'}(r_j'){Y_{\ell'}^{m'}}^*(\hat r_j') 
     \over \sin\delta^j_{\ell'}} \label{Green}
\end{eqnarray}
where $\kappa=\sqrt{z}$,
$\delta_\ell^i$ is the (complex) phase-shift for potential $V^i(r)$,
$t^i_\ell=\sin\delta^i_\ell\exp i\delta^i_\ell$,
$R^i_\ell(r)$ is the regular solution of the radial Schr\"odinger equation
for potential $V^i(r)$ that matches smoothly to $\cos\delta^i_\ell
j_\ell(\kappa r)-\sin\delta^i_\ell n_\ell(\kappa r)$ at the muffin-tin
radius $\rho_i$, 
$H^i_\ell(r)$ is the irregular solution of the radial Schr\"odinger equation
for potential $V^i(r)$ that matches smoothly to $h^+_\ell(\kappa r)$ at the 
muffin-tin radius. Finally the multiple-scattering matrix is
$\tau={[T_a^{-1}-\kappa H]}^{-1}$, where
${(T_a)}^{ij}_{\ell m\ell'm'}=
-(t^i_\ell/\kappa)\delta_{i,j}\delta_{\ell,\ell'} \delta_{m,m'}$ and 
${H}^{ij}_{\ell m\ell'm'}=-4\pi i\sum_{\lambda\mu}i^{\ell+\lambda-\ell'}
C^{\ell'm'}_{\ell m\lambda\mu}h^+_\lambda(\kappa R_{ij})
Y^\mu_\lambda(\hat R_{ij})$.  In the last expression, the Hankel
function $h^+_\lambda$ is defined as the function $h^{(1)}_\lambda$ of 
Ref.~\onlinecite{AbSt}. This straightforward extension of the real energy case
is possible because wave functions and phase-shifts are analytical functions
of $\kappa$. \cite{Taylor}

We present here the numerical
methods that were used to calculate the Green function and the 
ground state. First, we describe how the radial wave functions were obtained,
then we show how the Green function singularities were avoided,
 and finally
we explain how the multiple-scattering matrix was calculated.

\subsection{Radial wave functions \label{RadialSection}}
The core level energy and wavefunction were obtained from the Dirac
SCF program of Desclaux.\cite{Desclaux}

The photoelectron wave functions were obtained by adapting the method of
Ref.~\onlinecite{Koures}
to complex energies. A non-normalized regular wave function is written as
$\underline{R}_\ell(r)=r^\ell f_\ell(r)$ with $f_\ell(r)=f_r(r)+if_i(r)$.
For the complex energy $E=E_r+iE_i$ the radial Schr\"odinger equation
is :
\begin{eqnarray}
f_r''&=&(V(r)-E_r)f_r-{2(\ell+1)\over r}f_r'+E_if_i\nonumber\\
f_i''&=&(V(r)-E_r)f_i-{2(\ell+1)\over r}f_i'-E_if_r,
\end{eqnarray}
with the boundary conditions $f_r(0)=1$, $f_r'(0)=-Z/(l+1)$,
$f_i(0)=0$, $f_i'(0)=0$. Where $f'$ and $f''$ are, respectively, the first
and second radial derivatives of $f$. This system of second order differential
equations is transformed into a system of four first order differential 
equations which is solved by a forward step-adaptative fourth-order 
Runge-Kutta method. \cite{NumRec} 

The phaseshifts $\delta_\ell$ are deduced from the Wronskian equation :
\begin{equation}
\exp{(2i\delta_\ell)}=-{h^-_\ell(\kappa\rho)\underline{R}_{\ell}'(\rho)
-\kappa {h^-_\ell}'(\kappa\rho)\underline{R}_{\ell}(\rho)\over
h^+_\ell(\kappa\rho)\underline{R}_{\ell}'(\rho)
-\kappa {h^+_\ell}'(\kappa\rho)\underline{R}_{\ell}(\rho)}.
\end{equation}

The normalized radial wave functions $R^i_\ell(r)$ are obtained from
$\underline{R}^i_\ell(r)$ and the phaseshifts.

The irregular wave function is written as
$H_\ell=r^{-(\ell+1)} g_\ell(r)$ with $g_\ell(r)=g_r(r)+ig_i(r)$
and obtained by backward step-adaptative
fourth-order Runge-Kutta method \cite{NumRec} from the boundary
conditions $H_\ell(\rho)=h^+_\ell(\kappa\rho)$,
${H_\ell}'(\rho)=\kappa {h^+_\ell}'(\kappa\rho)$.

\subsection{Multiple-scattering matrix and singularities in the
 Green function}
\label{GFSing}

Formula (62) of Ref.~\onlinecite{Hikam} could not be directly 
applied because of possible infinite normalization factors arising from
the matching of the wave function at the muffin-tin radius to a sum of 
Bessel functions. \cite{Erreur}
With the present normalization at the muffin-tin radius, the radial
wave functions cannot be singular, but 
$\sin\delta^i_\ell$ terms arise in the
denominator of the Green function (see Eq.~\ref{Green}). Therefore, we must 
examine 
the behavior of the scattering matrix when $\sin\delta^i_\ell$ becomes small.
\cite{HikamTh}
We have
\begin{equation}
\tau=T_a{[1-\kappa H T_a]}^{-1}=
T_a+\kappa T_a{[1-\kappa H T_a]}^{-1}HT_a.
\end{equation}
Hence, 
\begin{equation}
\tau^{ij}_{\ell m\ell'm'}=-{1\over\kappa}\sin\delta^i_\ell e^{i\delta^i_\ell}
\left[
\delta_{i,j}\delta_{\ell,\ell'}\delta_{mm'}+
{\left({[1-\kappa H T_a]}^{-1}H\right)}^{ij}_{\ell m\ell'm'}
\sin\delta^j_{\ell'} e^{i\delta^j_{\ell'}}\right],
\end{equation}
and we see that $\tau^{ii}_{\ell m\ell m}/\sin^2\delta^i_\ell$ is
generally singular at $\sin\delta^i_\ell=0$ because of the first term.
These singularities 
did not appear in previous calculations of x-ray absorption cross
section because ${\rm Im}(\tau)$ was used, which is regular since
an additional $\sin\delta^i_\ell$ factor comes from $e^{i\delta^i_\ell}$.
To avoid these singularities, one can use 
$(\tau^{ii}_{\ell m\ell m}+t^i_\ell/\kappa)/\sin^2\delta^i_\ell$, which
is smooth. Therefore, all terms of the Green function (\ref{Green})
are regular.

To avoid the singularities discussed in the previous section, and 
to use a more symmetric equation, we use a polar decomposition 
technique \cite{Reed} and
define a modified multiple scattering matrix $\check{\tau}$ by
$\tau=T_a+\sqrt{T_a}\check{\tau}\sqrt{T_a}$. Numerically, $\check{\tau}$ is
obtained from the equation :
\begin{equation}
\check{\tau}={[1-\kappa \sqrt{T_a}H\sqrt{T_a}]}^{-1}-1.
\end{equation}

$\check{\tau}$ is a regularized multiple-scattering matrix that 
describes the effect,
on each atom, of the rest of the cluster ($\check{\tau}$ is zero
for a cluster of one atom).
The matrix elements of $[1-\kappa \sqrt{T_a}H\sqrt{T_a}]$ are calculated
from the phaseshifts and from the efficient algorithm for the calculation 
of $H$ that was proposed by Rehr and Albers
\cite{RehrAlbers} and tested against alternative methods.\cite{FatimaXAFS8}
The form
${[1-\kappa \sqrt{T_a}H\sqrt{T_a}]}^{-1}$ combines the advantages
of the two standard equations $\tau={[{T_a}^{-1}-\kappa H]}^{-1}$ 
and $\tau=T_a{[1-\kappa HT_a]}^{-1}$. As in the first equation, a
symmetric matrix is inverted (when real spherical harmonics are 
used), so that fast inversion algorithms
for symmetric matrices can be used.\cite{inversion} 
As in the second equation, it is
regular when ${(T_a)}^{ij}_{\ell m\ell'm'}=0$, so that spurious
singularities of $T_a^{-1}$ are very simply avoided. \cite{Case}
Moreover, asymptotic analysis \cite{Boston}
shows that it is well behaved when $\ell$ is large and/or $\kappa$ is
small, whereas none of the two standard equations is. The matrix is
inverted using a standard LU decomposition technique. \cite{NumRec}

\subsection{Cross-sections}
\label{crosssection}

The detailed expression of the absorption and XMCD cross-sections
for a powder relies on the orientational averaging technique 
used in Ref.~\onlinecite{Hikam}, and more calculation steps
are given in Ref.~\onlinecite{HikamTh}.
As in Eq.~(\ref{FermiInt}), all cross sections are written
as the sum of a Green function term and an integral.
With our modified multiple-scattering matrix, the ``Green function"
part of the K-edge absorption cross sections at energy
$E=\hbar\omega+E_i$ with a HWHM $\Gamma$ becomes:

\begin{equation}
\sigma_0=\sum_s{\rm Im}\left[{\tilde\sigma}_{0a}^s(E+i\Gamma)
+{\tilde\sigma}_{0n}^s(E+i\Gamma)\right]
\end{equation}
where the sum over $s$ is the sum over the two spin states (i.e. the
sum of the cross-sections calculated for potentials $V^\uparrow$ and
$V^\downarrow$). For
each spin state one defines
${\tilde\sigma}_{0a}(z)=(4\pi\alpha_0/3)(z-E_i)
i\sqrt{z}\exp[i\delta^0_1(z)]D^H(z)$, which
is the atomic contribution to x-ray absorption, and

\begin{equation}
{\tilde\sigma}_{0n}(z)=(4\pi\alpha_0/3)
(z-E_i)\sqrt{z}D^2(z)\exp[i\delta^0_1(z)]
{\check\tau^{00}(11,00;z)\over \sqrt{3}\sin\delta^0_1(z)} \label{sigma0n}
\end{equation}
which describes the influence of the neighbors ($i=0$ denotes
the absorbing site).
We recall that\cite{Hikam}
$\check\tau^{0j}(1\ell,a\alpha;z)=\sum_{\mu m}{(-1)^{(1-\mu)}
(1{\rm -}\mu\ell m|a\alpha)\check\tau^{0j}_{1\mu\ell m}(z)}$.

The radial integrals are $D(z)=\int_0^\infty{r^3dr\phi_0(r)R^0_1(r;z)}$,
where $\phi_0(r)$ is the large component of the core hole wave function and 
$D^H(z)=\int_0^\infty{r^3dr\phi_0(r)F(r;z)}$ where $F(r,z)$ is 
an auxiliary function defined as
$F(r;z)=\int_0^\infty{r'^3dr'\phi_0(r')R^0_1(r_<;z)H^0_1(r_>;z)}$
($R^i_\ell$ and $H^i_\ell$ are defined in section \ref{RadialSection}).

The ``Green function" part of the magnetic circular dichroism cross-section 
is written:

\begin{eqnarray}
\sigma_{MCD}={\rm Im}\Big[\sum_s((-1)^{(s-1/2)}
({\tilde\sigma}_{1a}^s(E+i\Gamma)
+{\tilde\sigma}_{1l}^s(E+i\Gamma)
+{\tilde\sigma}_{1n}^s(E+i\Gamma))\Big]
\end{eqnarray}
where ${\tilde\sigma}_{1a}(z)=(4\pi\alpha_0/3)(z-E_i)z
\exp[2i\delta^0_1(z)]M^{HH}(z)$
describes the purely atomic contribution to XMCD (the Fano effect),

\begin{equation}
{\tilde\sigma}_{1l}(z)=-(4\pi\alpha_0/3)(z-E_i)2iz
\exp[2i\delta^0_1(z)]D(z)M^{H}(z)
{\check\tau^{00}(11,00;z)\over \sqrt{3}\sin\delta^0_1(z)} \label{sigma1l}
\end{equation}

is the local contribution due to the spin-polarization of the
$p$-states on the absorbing site, and

\begin{eqnarray}
{\tilde\sigma}_{1n}(z)&=&(4\pi\alpha_0/3)(z-E_i)zD^2(z)\sum_{j\ell}
{(-1)^\ell\over 12}
\exp[i(\delta^0_1(z)+\delta^j_\ell(z))]\zeta^j_\ell(z)\nonumber\\
&\times&\sum_{a=|\ell-1|}^{\ell+1}[(\ell-a)(\ell+a+1)+2]\sum_\alpha
(-1)^{a-\alpha}
{\check\tau^{0j}(1\ell,a\alpha;z)\check\tau^{0j}
(1\ell,a{\rm -}\alpha;z)\over 
\sin\delta^0_1(z)\sin\delta^j_\ell(z)}
\label{sigma1n}
\end{eqnarray}
describes the contribution to XMCD due to the spin-orbit scattering
of the photoelectron by the neighbors and the absorber.
The radial matrix elements are
$M^H(z)=\int_0^\infty{r^2dr\xi(r)R^0_1(r;z)F(r;z)}$ ,
$M^{HH}(z)=\int_0^\infty{r^2dr\xi(r)F^2(r;z)}$ and
$\zeta^j_\ell(z)=\int_0^\infty{r^2dr\xi^j(r){[R^j_\ell(r;z)]}^2}$, with
$\xi(r)$ as defined in Eq.~(\ref{SOEq}).

The analysis of \S\ref{GFSing} shows that
$\check\tau^{0j}(1\ell,a\alpha;z)$ can be divided by
$\sqrt{\sin\delta^0_1(z)\sin\delta^j_\ell(z)}$, and all terms
of the XMCD cross-section are now regular.

For each term, the presence of the Fermi energy is taken into account
by calculating integrals along the line $e=E_F+it$ as in 
Eq.~(\ref{FermiInt}):
to all ${\rm Im}[\sigma(E+i\Gamma)]$ terms we add the integral:

\begin{equation}
{\Gamma\over\pi}\int_{0}^{\infty}dt
{\rm Re}\left[{\sigma(E_F+it)\over (E_F+it-E)^2+\Gamma^2}\right].
\end{equation}

\section{Results}

The converged potentials used in the multiple scattering formalism
were obtained by means of  an all-electron
self-consistent, scalar-relativistic and spin-polarized  linear-muffin-tin
orbital (LMTO) method. \cite{andersen}
We used the exchange-correlation potential and energy in
the von Barth and Hedin approximation. \cite{bart72} For the
Brillouin zone integration of the density of states
 we used the tetrahedron method  with about 300 ${\bf k}$
points in the irreducible part of the Brillouin zone.
\cite{jepsen72}  To simulate the effect of the core hole, we treated
the excited atom as a single impurity in a lattice using a supercell
geometry.
We have used increasingly larger supercells to ensure the convergence
of the magnetic moment and the density of
states of the impurity site. The final calculations were done for
a simple cubic lattice of 16 atoms per unit cell; the lattice parameter
being
$a = 2 a_0$. The results for the magnetic moment and the density of states
are
close to those of the supercell of 4 atoms. To use larger clusters in the
multiple-scattering formalism we had to assume that the potentials for
distant shells are bulk like.

\subsection{Near edge region}

Fig.~\ref{Febcc} a and b show the results we obtained for a converged
cluster of 259 atoms of bcc-iron (diameter 2.0~nm). The convergence
was investigated by checking that the spectral shape 
becomes stable with respect to the cluster diameter and by comparing 
to the results obtained with a cluster 
of 821 atoms (diameter 2.9~nm) on a wider energy mesh.
The potentials for the initial state (without core hole) and 
for the final state (with core hole) were obtained by a self-consistent 
super-cell calculation as indicated above. We used touching muffin-tin
spheres without overlap.
Because of the core hole width ($\Gamma = 2$~eV), absorption is possible
at energies lower than the Fermi level.
The effect of the Fermi level is clear, especially for the XMCD spectrum.
With a Fermi level $E_F = 2$~eV, the first positive structure disappears
whereas
with $E_F = -2$~eV, it is too large. Therefore, the size of the 
first positive 
peak depends strongly on the position of the Fermi level. 

The absorption and XMCD cross-sections are now smooth and the divergences
due to the cluster size and the relativistic corrections have disappeared.
We see also that the spectra are
continuous at the Fermi energy, although they are given by two
different formulas for $E>E_F$ and $E<E_F$. This corroborates the
fact that our treatment of the Fermi level is numerically sound.

Figure~\ref{Experiment} compares our calculation with experiment for
the edge region. All experimental and theoretical spectra were
normalized so that the absorption edge jump is 1.
The normal spectrum is not well reproduced.
The agreement for the XMCD spectrum is better, although the excellent
agreement in the intensity of the first two peaks is fortuitous,
because the degree of circular polarization was not one for the
experimental spectrum.
For the calculated
spectrum, various contributions are presented. The full thick line is
the total contribution, and the full thin line is the local contribution
($\tilde\sigma_{1l}$ of Eq.~(\ref{sigma1l})). 
In Fig.~\ref{sigman}, $\tilde\sigma_{1n}$ as written in
Eq.(\ref{sigma1n}) is expanded into the contribution of each $\ell$
($\ell$=1,2,3) and $j$=0,1,2 (absorbing atom, first shell, second
shell). For later use, the results are given without taking Fermi
energy into account. No term $\ell$=0 exists because $\ell$=0 gives
no spin-orbit. For the absorbing site, the terms with even $\ell$ 
are zero because of symmetry.
>From Fig.~\ref{sigman}, it can be observed that
the first positive peak (peaking at 1~eV)
is not due to the local term (\ref{sigma1l}),
but to the spin-polarization
of the $d$-states of the neighbors (mainly the term corresponding to
$\ell$=2 and $j$=1 in Eq.~(\ref{sigma1n})).
This is related to the weak ferromagnetic nature of bcc-Fe,
\cite{Pizzini94} and
was already observed on a smaller cluster \cite{HikamTh} and
in tight-binding calculations. \cite{Igarashi94,Igarashi96}
Both absorption and XMCD calculated structures are too large from
20~eV above the edge.
This reduction, which is very common in multiple-scattering calculations,
is probably due to photoelectrons
that experience inelastic interaction with the metal. This inelastic
effect comes into play above the plasmon energy (about 10~eV)
and can be included in our calculation through 
an energy-dependent $\Gamma$.\cite{Sainctavit,Tyson2}
The main failure of the XMCD calculation is the presence of a large
second positive peak which is absent in the experiment. A similar
peak can be observed in fully-relativistic \cite{Ebert88} and
tight-binding calculations,\cite{Igarashi94,Igarashi96}
although a direct comparison is difficult because of the different
normalization used.

\subsection{Core hole}
Two modifications of the core hole were tested. In the first one,
the non relativistic (Schr\"odinger) equation was solved for the
core hole. The core hole energy was found different, but the
$1s$ wavefunction was quite similar to the relativistic one,
and the normal spectrum was a bit smaller (because the normalization
of the relativistic wavefunction
includes the small component, which does not contribute to the
spectrum) but the XMCD spectrum was not distinguishable from the 
relativistic result. Similarly, core hole exchange splitting is
negligible since the difference between the core states
obtained with up and down potentials did not yield noticeable effects.

The second test was conducted to test the 
influence of the core hole on the spectra.  Fig.~\ref{hole} 
shows the normal and XMCD spectra with and without core hole
and with a Fermi energy $E_F$=0~eV).
The normal spectra are quite similar, except for an overall amplitude
factor. The XMCD spectra are more different, although the position of
the structure does not move. Comparison with experiment does not
enable us do decide for a model.

\subsection{SPEXAFS}
In XMCD at the K-edge, spin-orbit acts directly on the photoelectron,
and it is interesting to know whether the high-energy part of the
spectrum is related to the spin polarization of the $p$-states of
the absorbing site. Using the methods developed in 
Ref.~\onlinecite{KikiRino}, it is possible to show that, in the
single scattering approximation, only the $p$-projected term
${\tilde\sigma}_{1l}^\uparrow-{\tilde\sigma}_{1l}^\downarrow$ survives
at high energy.
Figure~\ref{EXAFS}(b) shows the spectrum obtained with a cluster
of 51 atoms, calculating all angular momenta up to $\ell$=8, and
multiplied by 1/3.
The cluster is too small to be realistic in the edge
region, but above 30~eV, the overall agreement is correct. 
The peak at 110~eV in the theoretical
spectrum should be broadened, and the peak at 60~eV in the experimental
spectrum is probably due to multielectronic effects, which are known
to be strong in XMCD spectra at that energy. \cite{Dartyge} 
The thin line represents
the contribution of the $p$-projected states, which is seen to be dominant
at high energy. 
Moreover, the calculated XMCD spectrum
is in phase with the calculated EXAFS spectrum. This phase relation
between dichroic and normal spectra
was observed experimentally by Pizzini and coll. \cite{Pizzini94} 

It is sometimes assumed that XMCD reflects the spin-polarization of $p$
states projected on the absorbing atom ($\rho^\uparrow-\rho^\downarrow$).
Fig.~\ref{pDOS} shows that the absorption spectrum is indeed quite
similar to the $p$ density of states, but the XMCD spectrum is 
quite different from the spin polarization of the $p$ density of 
states. In fact, the spin-polarization is very similar to the
derivative of the density of states. In other words, the rigid band
model becomes correct at high energy (a somewhat surprising result), 
and the band splitting is about 1~eV.

>From this result, a very simple approximate expression can be 
derived for XMCD. Assuming a rigid band model, we can consider
that the up and down bands are exchange-splitted by the energy
$\Delta E$. Moreover, neglecting the non-diagonal terms of the
spin-orbit operator, one can consider that the $m=\pm 1$ components
of the $p$-band are split by spin-orbit coupling $\zeta 
{\bf \ell . s}$. Therefore, for transitions towards $\ell=1,m=1$
final states (left-circularly polarized x-rays), we have
\begin{eqnarray}
\sigma^{+\uparrow}&=&\sigma(E+\Delta E/2-\zeta)\\
\sigma^{+\downarrow}&=&\sigma(E-\Delta E/2+\zeta)\\
\sigma^+&\simeq&2\sigma(E)+(\Delta E/2-\zeta)^2 d^2\sigma/dE^2
\end{eqnarray}
and for transitions towards $\ell=1,m=-1$ final states
(right-circularly polarized x-rays):
\begin{eqnarray}
\sigma^{-\uparrow}&=&\sigma(E+\Delta E/2+\zeta)\\
\sigma^{-\downarrow}&=&\sigma(E-\Delta E/2-\zeta)\\
\sigma^-&\simeq&2\sigma(E)+(\Delta E/2+\zeta)^2 d^2\sigma/dE^2
\end{eqnarray}
therefore, XMCD becomes
\begin{equation}
\sigma^+-\sigma^-\simeq-2\Delta E \zeta d^2\sigma/dE^2.
\end{equation}
Since $\zeta$ is fairly constant at high energy, the
exchange-splitting energy $\Delta E$ can be deduced from
experimental spectra. Figure~\ref{Deriveeseconde} shows
XMCD together with the second derivative of the normal
spectrum (multiplied by 10). Because the model is very
crude, the agreement is not perfect, but good enough to
say that the image deduced from the rigid-band model is correct.
The presence of this second derivative explains also the
phase relation between EXAFS and XMCD (considering EXAFS as a sum
of sines). A more elaborate interpretation comes from considering
the expression for $\tilde\sigma_{0n}$ (Eq.~(\ref{sigma0n})) and
$\tilde\sigma_{1l}$ (Eq.~(\ref{sigma1l})). Both expressions involve
the multiple-scattering matrix $\check\tau^{00}(11,00;z)$,
but $\tilde\sigma_{1l}$ contains an additional factor $i$ that makes
it proportional to the real part of $\check\tau^{00}(11,00;z)$,
whereas the EXAFS cross section is proportional to its imaginary part
(it was checked that the other factors do not intervene much in the phase
at high energy). Since real and imaginary part of the Green function
are related by Kramers-Kronig theorem, which transforms sine functions
into cosine functions, and because dichroic effect is due to the
difference between spin up and spin down $\tilde\sigma_{1l}$, it is 
proportional to the derivative of the real part of 
$\check\tau^{00}(11,00;z)$ which, because of an additional minus sign,
is in phase with the imaginary part of the multiple-scattering matrix
(to see this, consider that, in the EXAFS regime, the Green function
can be approximate by $\exp(ikR+\phi)$, the imaginary part is a sine
function - the EXAFS formula - and the derivative of the real
part is a sine function as well). Of course, this correspondence
between XMCD and EXAFS is not exact, and much interesting information
comes from the difference.

\subsection{Multiple-scattering expansion}
The program that we used to calculate the XMCD spectrum of iron
is relatively heavy and slow. To make the analysis of XMCD spectra
as easy as that of normal absorption spectra, it is necessary to
investigate the validity of the multiple-scattering expansion
which is used by much faster programs, such as FEFF. \cite{Rehr}
To do this, we compare the results obtained by the full inversion
of $(1-\kappa T_aH)^{-1}T_a$ with the single scattering expansion
($T_a+\kappa T_aHT_a+\kappa^2 T_aHT_aHT_a$) and the double-scattering
expansion (previous term plus $\kappa^3{(T_aH)}^3T_a$).
This comparison is shown in Fig.~\ref{MultipleScattering}.
The overall agreement is correct but not excellent, probably because
of the shadowing effect which is large in bcc structures.
Notice that, in principle, the zeroth-order scattering
($T_a+\kappa T_aHT_a$) also contributes due to the term 
$\tilde\sigma_{1n}$. However, this contribution is very small
at high energy. Higher order terms are considered in 
Ref.~\onlinecite{brouderHerrsching}.

\section{Conclusion}
Since the first experimental XMCD spectra,\cite{Schutz87}
a number of calculations of XMCD K-edge spectra have been carried out,
using different methods. A fully relativistic KKR method
was used for 
bcc-Fe,\cite{Ebert88,Ebert88b,Ebert88c,Ebert91c,Stahler93,Gotsis94}
hcp-Co,\cite{Ebert91c,Strange90} Fe-Co alloys,
\cite{Ebert92b,Ebert93b,Ebert93,Gotsis94b,Strange95,Ebert96}
and cfc-Ni.\cite{Stahler93} A relativistic LMTO-ASA calculation of
the K-edge XMCD spectrum of Fe in GdFe$_{2}$ was carried out
in Ref.~\onlinecite{Lang94}, a molecular orbital approach
was used for Fe in tetrahedral and octahedral 
environments,\cite{Harada94} 
a multiplet approach
for Ni in a molecular magnet \cite{Verdaguer95} and
a tight-binding method
for metallic Fe and Ni \cite{Igarashi94,Igarashi96} and
Co, \cite{Igarashi96} where XMCD was related to the projection
of the orbital momentum along the x-ray direction.
All these calculations
were restricted to a narrow energy range around the edge. 
We have developed in this paper a multiple-scattering approach
which allows for the calculation of extended structure,
and includes the core hole without additional effort.

We have presented a solution of the convergence difficulties associated to 
the one-electron
calculation of physical properties related to spin-orbit interaction. 
A recourse to Green functions
with complex energy argument led us to smooth absorption and XMCD 
cross-sections. The presence of the Fermi level was accounted for through
a complex plane integration which was found much more stable than on
the real line. This technique can be used to calculate other spin-orbit
influenced properties, such as anisotropy energy or spin-dependent 
spectroscopies.

Robust and accurate numerical methods were proposed to evaluate the Green
function in the whole complex plane, and the smooth behavior of the
cluster Green function for large imaginary energies was explained.

We saw that the EXAFS part of XMCD at the K-edge is simply
connected to the spin-polarization of the $p$-states on the absorbing
site. The high-energy part is therefore simpler to interpret than the
near-edge part, where spin-orbit interactions with the neighbors
gives strong contributions.

Our first application is encouraging, and further comparison with experiment,
including the relation between magnetic and non-magnetic fine structure,
\cite{Pizzini94} will be presented in a forthcoming publication.
Multiple-scattering calculation of XMCD at the 
L$_{\mbox{\scriptsize{II,III}}}$-edges of Gd are presented in
Refs.~\onlinecite{brouderHerrsching} and
\onlinecite{brouderMittelwihr}.

As observed by Ankudinov and Rehr, \cite{Ankudinov} XMCD in the 
x-ray range provides
a very good test of effective potentials representing the exchange
interaction of the photoelectron with matter. An alternative
exchange potential was proposed by Zhogov et al. \cite{Zhogov95}
in their treatment of magnetic EXAFS.
In the present study,
we have used the potential provided by the ground and relaxed-excited
states, but did not include any specific exchange potential.

\acknowledgments
We thank E. Dartyge, S. Pizzini, Ch. Giorgetti, F. Baudelet
and A. Fontaine for providing us with the experimental spectra of bcc-Fe.
We gratefully acknowledge the encouragement of G. B. White, who gave us
detailed explanations of the mathematical framework. We thank
also  F. Gesztesy and B. Thaller for helpful discussions  
on the mathematical difficulties of the  F-W Hamiltonian.
We are very grateful to S. Hagenah, who suggested the contour integration
to  account for the Fermi energy, to J.J. Rehr for communication
of his XMCD results prior to publication,  to Ph. Sainctavit for his
thorough reading of the manuscript, and to J. W. Wilkins for 
interesting discussions.
One of us (Ch.B.) would like to thank K. H. Bennemann for his hospitality.
This work was partly supported by NATO grant 9C93FR, by C.N.R.S.
and by C.N.R.S-NSF cooperative research program (Grant n$^{\rm o}$
INT-9314455).  M. A. acknowledges  partial support provided by the Department
of Energy (DOE) - Basic Energy Sciences, Division of Materials Sciences, and
Supercomputer time provided by the Ohio Supercomputer.

\appendix
\section*{Symmetrized basis}
One of the advantages of the semi-relativistic limit of the Dirac
Green function is that one is allowed to use the full local point
symmetry of the absorbing atom without taking spin direction into
account, since the spin and space functions are not coupled.
In this section, we show how symmetrized bases were implemented to
reduce the size of the problem.

For each spin state, the cluster potential is written as 
$V({\bf r})=\sum_i{V^i(r_i)}$
($i$ runs over atomic sites),
and let $P_a$ represent the action of the symmetry operation $a$ 
on the cluster. If $a$ belongs to the local point symmetry 
group ${\cal G}$ of the absorbing atom ($i=0$), 
then $P_a[V({\bf r})]=V({\bf r})$.
Therefore, the Green function $G(z)=(z-H_0-V)^{-1}$ is also invariant
for operations of ${\cal G}$. This can be used to reduce the size of the
multiple-scattering matrices. \cite{Diamond,LF}

In practice, we set up a symmetrized basis by the following procedure.
For an irreducible representation (irrep) $\alpha$ of the symmetry group,
we take a matrix realization $\Gamma^{(\alpha)}$ of $\alpha$, and we
define the (pseudo-)projector:
\begin{equation}
P^{(\alpha)\ell}_{jj_0}={d_\alpha\over g}\sum_{a\in \cal G} 
  \Gamma^*_{jj_0}(a) P^\ell_{}(a)
\end{equation}
where the cluster symmetry operator $P^\ell(a)$ has matrix elements
in the $|n\ell m\rangle$ basis :
\begin{equation}
  [P^\ell(a)]_{n'm'nm}= D^\ell_{m'm}(a) \delta_{n',a(n)}.
\end{equation}
Here, $g$ is the number of elements of the symmetry group 
${\cal G}$, $d_\alpha$
is the dimension of irrep $\alpha$,
$D^\ell_{m'm}(a)$ is the (eventually improper) Wigner rotation
matrix corresponding to the symmetry operation $a$ of the point
symmetry group ${\cal G}$. \cite{Biedenharn}
We choose a column $j_0$ of the matrix realization ($j_0=1$) and,
for each $|n\ell m\rangle$ representing a spherical harmonics
$Y_\ell^m$ attached to site $n$, we calculate the projection
$P^{(\alpha)\ell}_{j_0j_0}|n\ell m\rangle$. 
Let $p$ be the set of atoms that can be obtained from a given atom 
by operations of ${\cal G}$.
For each $\ell$ and each $p$ one can generate a vector space 
$E(\alpha p\ell j_0)$ by calculating
all the projections of $|n\ell m\rangle$, for $n\in p$ and $m=-\ell,\dots,\ell$.
By singular value decomposition, \cite{NumRec} an orthonormal basis of 
$E(\alpha p\ell j_0)$
is obtained, which is denoted by $|\alpha p\ell j_0 s\rangle$,
where $s=1,...,\dim(E(\alpha p\ell j_0))$. This symmetrized basis can
be written as a function of the initial basis kets :
\begin{equation}
|\alpha p\ell j_0 s\rangle=\sum_{n\in p \atop m=-\ell,...,\ell} 
            \langle n\ell m|\alpha p\ell j_0 s\rangle |n\ell m\rangle.
\end{equation}

The partners of 
$|\alpha p\ell j_0 s\rangle$ are obtained by 
\begin{equation}
|\alpha p\ell j s\rangle=P^{(\alpha)\ell}_{jj_0}|\alpha p\ell j_0 s\rangle
\end{equation}

The symmetrized basis is then used to simplify the matrix inversion
yielding the relevant multiple scattering matrix elements $\tau^{0i}_{L L'}$.
It can be shown that
\begin{equation}
(P^\ell(a) H)_{nlm\,n'l'm'}=(HP^{\ell'}(a))_{nlm\,n'l'm'}
\end{equation}
Therefore, 
\begin{eqnarray}
\langle \alpha p \ell j s|H|\alpha' p' \ell' j' s'\rangle&=&
\langle \alpha p \ell j_0 s|P^{(\alpha)\ell}_{j_0j}H
   P^{(\alpha')\ell'}_{j'j_0}
   |\alpha' p' \ell' j_0' s'\rangle\\
&=&\langle \alpha p \ell j_0 s|P^{(\alpha)\ell}_{j_0j}
   P^{(\alpha')\ell}_{j'j_0}
   H|\alpha' p' \ell' j_0 s'\rangle\\
&=&\delta_{\alpha,\alpha'}\delta_{j,j'}
  \langle \alpha p \ell j_0 s|H|\alpha p' \ell' j_0 s'\rangle.
\end{eqnarray}
Similarly, the atomic scattering matrix is diagonal:
\begin{equation}
\langle \alpha p \ell j s|T_a^{-1}|\alpha' p' \ell' j' s'\rangle=
-{\kappa\over t^p_\ell}\,
\delta_{\ell,\ell'} \delta_{p,p'}\delta_{\alpha,\alpha'}\delta_{j,j'}
\delta_{s,s'}
\end{equation}
where $t^p_\ell=\sin\delta^p_\ell\exp(i\delta^p_\ell)$ and $\delta^p_\ell$
is the $\ell$-th phaseshift of the atoms of type $p$.
Because of the transformation properties of matrices $H$ and $T_a$,
the full multiple-scattering matrix $\tau={[T_a^{-1}-\kappa H]}^{-1}$ 
is diagonal in $\alpha$ and $j$:
\begin{equation}
\langle \alpha p \ell j s|\tau|\alpha' p' \ell' j' s'\rangle=
\delta_{\alpha,\alpha'}\delta_{j,j'}
\tau^{(\alpha)}_{p\ell s\,p'\ell's'}.
\end{equation}
In other words, the multiple-scattering matrix can be inverted separately
for each irrep and it repeats itself $\dim(\Gamma^{(\alpha)})$ times in
each irrep. For the example of a cluster of 259 Fe atoms, only the
$T_{1u}$ irrep is relevant, because of dipole selection rules, and
the matrix to be inverted has dimension 288 instead of 4144.
Once $\tau^{(\alpha)}_{p\ell s\,p'\ell's'}$ is obtained, the
matrix elements which are required for the calculation of XMCD at
the K-edge are
obtained through the basis change:
\begin{equation}
\tau^{0i}_{1 \epsilon \,\ell m}=\sum_{\alpha j s s'}
\langle 0 1 \epsilon|\alpha p_0 1 j s\rangle
\tau^{(\alpha)}_{p_0 1 s\,p_i\ell s'}
\langle \alpha p_i \ell j s'| i\ell m \rangle
\end{equation}
where $p_i$ is the class of atoms to which atom $i$ belongs.

An additional advantage of the use of a symmetrized basis is the fact
that the inverse of $\tau$ is more rapidly calculated. In practice, the
matrix elements 
$\langle \alpha p \ell j s|H|\alpha p' \ell' j s'\rangle$ must be
calculated from the matrix elements $H^{nn'}_{\ell m\ell'm'}$ and the
basis change matrix by
\begin{equation}
\langle \alpha p \ell j_0 s|H|\alpha p' \ell' j_0 s'\rangle=
\sum_{n\in p \atop n'\in p'}\sum_{m m'}
\langle\alpha j_0\ell p s|n\ell m\rangle
H^{nn'}_{\ell m\ell'm'}
\langle n'\ell' m'|\alpha j_0\ell' p' s'\rangle.
\end{equation}
Instead of summing over all pairs $n\in p$ and $n' \in p'$, we can
choose a member $n_0$ of class $p$, and a set of members $n_1$ of class
$p'$ which span all possible non-equivalent neighbors of $n_0$ in 
class $p'$. In other words, for each pair $(n,n')$, there is an element
$a$ of symmetry group ${\cal G}$ and a representative $n_1$ such that 
$a(n)=n_0$ and $a(n')=n_1$.
Therefore
\begin{equation}
H^{nn'}_{\ell m\ell'm'}=\sum_{\mu,\mu'}
{D^\ell_{\mu m}}^*(a)
{D^\ell_{\mu' m'}}(a)
H^{n_0n_1}_{\ell \mu\ell'\mu'}.
\end{equation}

We have also
\begin{equation}
\sum_{m} D^\ell_{\mu m}(a) 
\langle n\ell m|\alpha j_0\ell p s\rangle=
\sum_{j}\Gamma^{\alpha}_{j_0 j}(a^{-1})
\langle n_0\ell \mu|\alpha j\ell p s\rangle,
\end{equation}
which yields, using the fact that the symmetrized matrix elements
of $H$ do not depend on $j_0$:
\begin{eqnarray}
&&\langle \alpha p \ell j_0 s|H|\alpha p' \ell' j_0 s'\rangle=
{1\over d_\alpha}\sum_{j,j',j''}\sum_{n_1,a}
\Gamma^{\alpha}_{j'' j}(a)
{\Gamma^{\alpha}_{j'' j'}}^*(a)\\
&\times&\sum_{m m'}
\langle\alpha j\ell p s|n_0\ell m\rangle
H^{n_0n_1}_{\ell m\ell'm'}
\langle n_1\ell' m'|\alpha j'\ell' p' s'\rangle.
\end{eqnarray}

>From the orthogonality theorem of group representations, \cite{LF}
the sum over $j''$ does not depend on $a$, and 
\begin{equation}
\langle \alpha p \ell j_0 s|H|\alpha p' \ell' j_0 s'\rangle=
\sum_{n_1,j}{|p|N_{n_0n_1}\over d_\alpha}
\langle\alpha j\ell p s|n_0\ell m\rangle
H^{n_0n_1}_{\ell m\ell'm'}
\langle n_1\ell' m'|\alpha j\ell' p' s'\rangle,
\end{equation}
where $|p|$ is the number of elements of class $p$, and $N_{n_0n_1}$
is the number of elements $n'$ of $p'$ such that $a(n_0)=n_0$ and
$a(n_1)=n'$, where $a\in {\cal G}$.
Therefore, the sum is reduced to pairs of inequivalent neighbors.

>From these symmetry considerations, one can show that, at the K-edge of
a transition metal in a cubic environment, the spin-polarization of the
$d$-shell of the absorbing atom cannot be measured, but the spin-polarization
of the $d$-shell of the neighbors can. This was observed in  the
multiple-scattering approach in
Refs.~\onlinecite{Pizzini94} and \onlinecite{HikamTh} and in 
the tight-binding approach in Ref.~\onlinecite{Igarashi94,Igarashi96}.
%

%
%
\begin{figure}
\caption{\label{FebccEi0} 
Calculated  K-edge absorption spectrum (thick line)  and  x-ray magnetic
circular dichroism (thin line) of a cluster of 259 atoms of bcc-Fe near
the singularity at the vicinity of Fermi level.}
\end{figure}

\begin{figure}
\caption{\label{poles} Pole structure of the integrand in the second term of
Eq.~(\protect\ref{EfInfty}).}
\end{figure}

\begin{figure}
\caption{\label{G(EF+it)} (a) : $\sum_s {\tilde \sigma}_{0a}^s+
{\tilde \sigma}_{0n}^s$ and (b) 
: $\sum_s (-1)^{s-1/2}{\tilde \sigma}_{1a}^s+
{\tilde \sigma}_{1l}^s+ {\tilde \sigma}_{1n}^s$
for complex energies
along the line $E_F+it$ ($E_F=0$). In each case, the corresponding atomic 
quantities 
$\sum_s {\tilde \sigma}_{0a}^s$ and
$\sum_s (-1)^{s-1/2}{\tilde \sigma}_{1a}^s$
are plotted, to show the convergence of the full Green
function to the atomic Green function.}
\end{figure}

\begin{figure}
\caption{\label{Febcc} 
Calculated (a) K-edge absorption spectrum and (b) x-ray magnetic circular 
dichroism (XMCD) of a cluster of 259 atoms of bcc-Fe with different values of 
Fermi level, $E_F=-\infty,-2,0,2$~eV.
For the XMCD signal, the size of the first positive structure 
depends strongly on the position of Fermi level, and disappears for $E_F$ 
above 2~eV. }
\end{figure}

\begin{figure}
\caption{\label{Experiment} 
Comparison  of the  calculated (a) K-edge absorption   
spectrum and (b) the x-ray magnetic circular dichroism (XMCD) of a cluster of 
259 atoms of bcc-Fe with the 
experimental results of Ref.~\protect\onlinecite{Pizzini94} 
The Fermi level is $E_F$=0~eV.}
\end{figure}

\begin{figure}
\caption{\label{sigman}
Contribution to $\tilde\sigma_{1n}$ of the absorbing site and the
first two shells (from top to bottom) and for $\ell$=1,2,3.
No Fermi energy was used.
The contribution of a given shell is the contribution of one
atom of the shell multiplied by the number of atoms in the shell.}
\end{figure}

\begin{figure}
\caption{\label{hole} 
Calculated (a) K-edge absorption spectrum and (b) x-ray magnetic circular 
dichroism (XMCD) of a cluster of 259 atoms of bcc-Fe with (thick line)
and without (thin line) core hole.}
\end{figure}

\begin{figure}
\caption{\label{EXAFS}  Comparaison of the calculated (a) K-edge absorption 
spectrum and (b)
x-ray magnetic circular dichroism (XMCD) of a cluster of 51 atoms of
bcc-Fe with the experimental results of Ref.~\protect\onlinecite{Pizzini94}.
The Fermi level is $E_F$=0~eV, the core-hole broadening is
$\Gamma=2$~eV, and the maximum scattering wave is $\ell_{\max}$=8.}
\end{figure}

\begin{figure}
\caption{\label{pDOS}  (a) Comparison of the calculated K-edge absorption 
spectrum with the density of $p$ states projected on the absorbing atom
and (b) comparison of the calculated XMCD with the spin polarization
of the density of $p$ states projected on the absorbing atom. The
derivative of the DOS is also presented, to illustrate the rigid band
picture that becomes valid at high energy.}
\end{figure}

\begin{figure}
\caption{\label{Deriveeseconde}  Comparison of the XMCD spectrum with
the second derivative of the absorption spectrum (multiplied by 10) 
for a cluster of 51 iron atoms.}
\end{figure}

\begin{figure}
\caption{\label{MultipleScattering}  (a) Comparison of the 
calculated K-edge absorption 
spectrum with the single and double scattering expansion (b) comparison 
of the calculated XMCD with the single and double scattering
expansion.}
\end{figure}


\begin{references}
\bibitem[*]{byline} Part of this work was carried out during a stay
  at the Institut f\"ur Theoretische Physik, Arnimallee 14,
  D-14195 Berlin.
\bibitem{Jordan} This can be shown 
  from the inequality $|G(z)|\le 1/d(z,\Sigma)$, where $d(z,\Sigma)$
  is the distance between $z$ and the spectrum of $H$. If $E_0$ is the
  energy of the 1s state, $d(z,\Sigma)\ge \max(E_0-\rm{Re}z,|\rm{Im}z|)$.
\bibitem{Erreur} and because the signs of the last two terms of 
  Eq.~(62) in Ref.~\onlinecite{Hikam} are wrong.
\end{references}
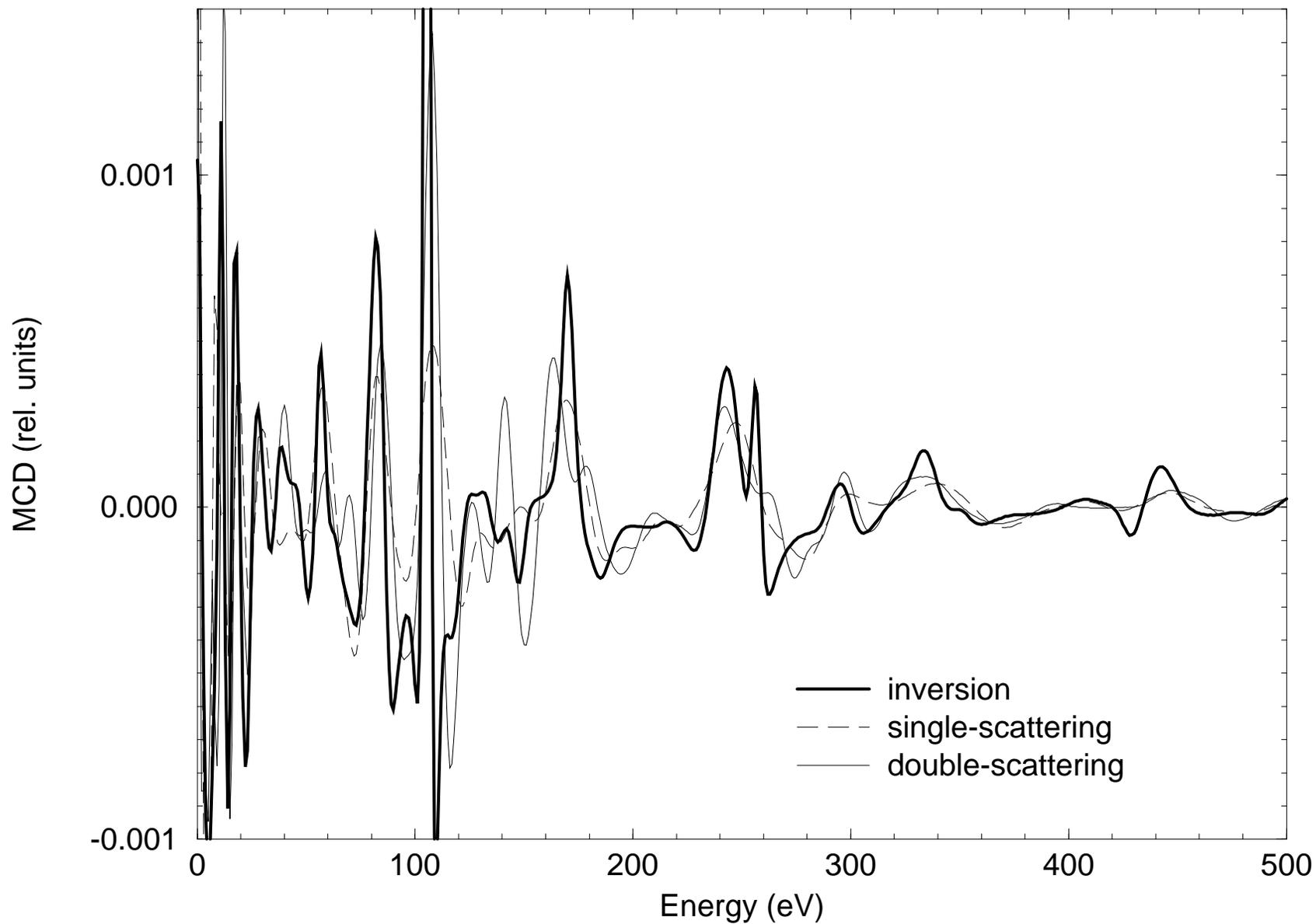
\end{document}